\documentclass[%
 aip,
 amsmath,amssymb,
 reprint,%
]{revtex4-1}
\usepackage{graphicx}
\usepackage{dcolumn}
\usepackage{bm}
\usepackage{comment}
\usepackage{xcolor}
\usepackage{hyperref}
\usepackage{amssymb}
\usepackage{float}
\usepackage{booktabs}
\usepackage{multirow} 
\usepackage{physics}

\usepackage[utf8]{inputenc}
\usepackage[T1]{fontenc}
\usepackage{mathptmx}
\usepackage{etoolbox}
\usepackage[version=4]{mhchem}
\usepackage{soul}
\usepackage{siunitx}
\usepackage{xparse}
\usepackage{xcolor}
\usepackage{pifont}
\definecolor{mygreen}{RGB}{0, 128, 0}
\definecolor{myred}{RGB}{128, 0, 0}
\usepackage{amssymb,bbding}
\usepackage{caption}
\makeatletter
\def\@email#1#2{%
 \endgroup
 \patchcmd{\titleblock@produce}
  {\frontmatter@RRAPformat}
  {\frontmatter@RRAPformat{\produce@RRAP{*#1\href{mailto:#2}{#2}}}\frontmatter@RRAPformat}
  {}{}
}%
\makeatother
\begin{document}

\preprint{AIP/123-QED}

\title[Microgram $\mathrm{BaCl}_2$ Ablation Targets for Trapped Ion Experiments]{Microgram $\mathrm{BaCl}_2$ Ablation Targets for Trapped Ion Experiments}
\author{Noah Greenberg}
 \altaffiliation[n2greenb@uwaterloo.ca]{}
\author{Akbar Jahangiri Jozani}
\author{Collin J. C. Epstein}
\author{Xinghe Tan}
\author{Rajibul Islam}
\author{Crystal Senko}
\affiliation{ 
Institute for Quantum Computing and Department of Physics and Astronomy, University of Waterloo, Waterloo, Ontario N2L 3G1, Canada
}%

\date{\today}

\begin{abstract}
Trapped ions for quantum information processing has been an area of intense study due to the extraordinarily high fidelity operations that have been reported experimentally. 
Specifically, barium trapped ions have been shown to have exceptional state-preparation and measurement (SPAM) fidelities. 
The $^{133}\mathrm{Ba}^+$ ($I = 1/2$) isotope in particular is a promising candidate for large-scale quantum computing experiments.
However, a major pitfall with this isotope is that it is radioactive and is thus generally used in microgram quantities to satisfy safety regulations. 
We describe a new method for creating microgram barium chloride ($\mathrm{BaCl}_2$) ablation targets for use in trapped ion experiments and compare our procedure to previous methods.
We outline two recipes for fabrication of ablation targets that increase the production of neutral atoms for isotope-selective loading of barium ions. 
We show that heat-treatment of the ablation targets greatly increases the consistency at which neutral atoms can be produced and we characterize the uniformity of these targets using trap-independent techniques such as energy dispersive x-ray spectroscopy (EDS) and neutral fluorescence collection. Our comparison between fabrication techniques and demonstration of consistent neutral fluorescence paves a path towards reliable loading of $^{133}\mathrm{Ba}^+$ in surface traps and opens opportunities for scalable quantum computing with this isotope.
\end{abstract}


\maketitle

\section{\label{sec:intro}Introduction}
Ion trapping is a leading platform for quantum simulation and quantum computing due to its scalability \cite{PhysRevA.89.022317,article1}, high-fidelity entangling gates \cite{PhysRevLett.117.060504,PhysRevLett.131.063001}, and long coherence times \cite{Wang2021-es}. 
A leading trapped ion species that has recently emerged due to its unrivaled state-preparation and measurement fidelities is barium, specifically the $^{137}\mathrm{Ba}^+$ and $^{133}\mathrm{Ba}^+$ isotopes\cite{PhysRevLett.129.130501,Christensen2020,PhysRevLett.119.100501}. 
Barium has many attractive atomic features such as visible wavelength transitions for easier optical engineering\cite{Binai_Motlagh_2023} and a metastable state for shelving\cite{low2023control,vizvary2023eliminating}.

The caveats to the $^{133}\mathrm{Ba}$ isotope are that it is typically sourced as a synthetic radioactive salt, either as $\mathrm{BaCl}_2$ or $\mathrm{Ba}(\mathrm{NO_3})_2$, and can only be used in quantities deemed safe by a governing regulatory body. 
For example, the Canadian Nuclear Safety Commission (CNSC) dictates that holders of a Basic Level license can work with up to $\sim 55.5$~MBq of the $^{133}\mathrm{Ba}$ isotope\cite{ba133limit}.
In this manuscript, we aim to create a recipe for the fabrication of reliable microgram quantity atomic sources for trapped ion experiments in order to simplify regulatory requirements, reduce worker exposure, and increase the duration that a worker can safely be near the radioactive source. 
We choose to work with $\mathrm{BaCl}_2$ over $\mathrm{Ba}(\mathrm{NO_3})_2$ since  $\mathrm{BaCl}_2$ has been used to previously trap the $^{133}\mathrm{Ba}$ isotope\cite{Christensen2020}.

Loading an ion trap reliably with microgram quantities of any isotope is difficult and so far $^{133}\mathrm{Ba^+}$ trapping has only been reported by a sole academic group\cite{Christensen}, despite there being significant interest from the research community \cite{PhysRevA.105.033102,mitsurface,huculconference,brennanconference}.
We present the first systematic study of microgram $\mathrm{BaCl}_2$ ablation targets for use in trapped ion experiments and fabricate targets that produce consistent and reliable neutral fluorescence\cite{greenberg2023trapping}. 
The longevity of the fabricated ablation targets as well as the consistency at which the targets produce measurable neutral fluorescence is investigated. 
Given that none of the properties we investigate are isotope-dependent, we use $^{138}\mathrm{Ba}$ as a proxy for $^{133}\mathrm{Ba}$ because $^{138}\mathrm{Ba}$ is naturally abundant and nonradioactive. 

\begin{figure}
\centering
\includegraphics[width = 0.9\linewidth]{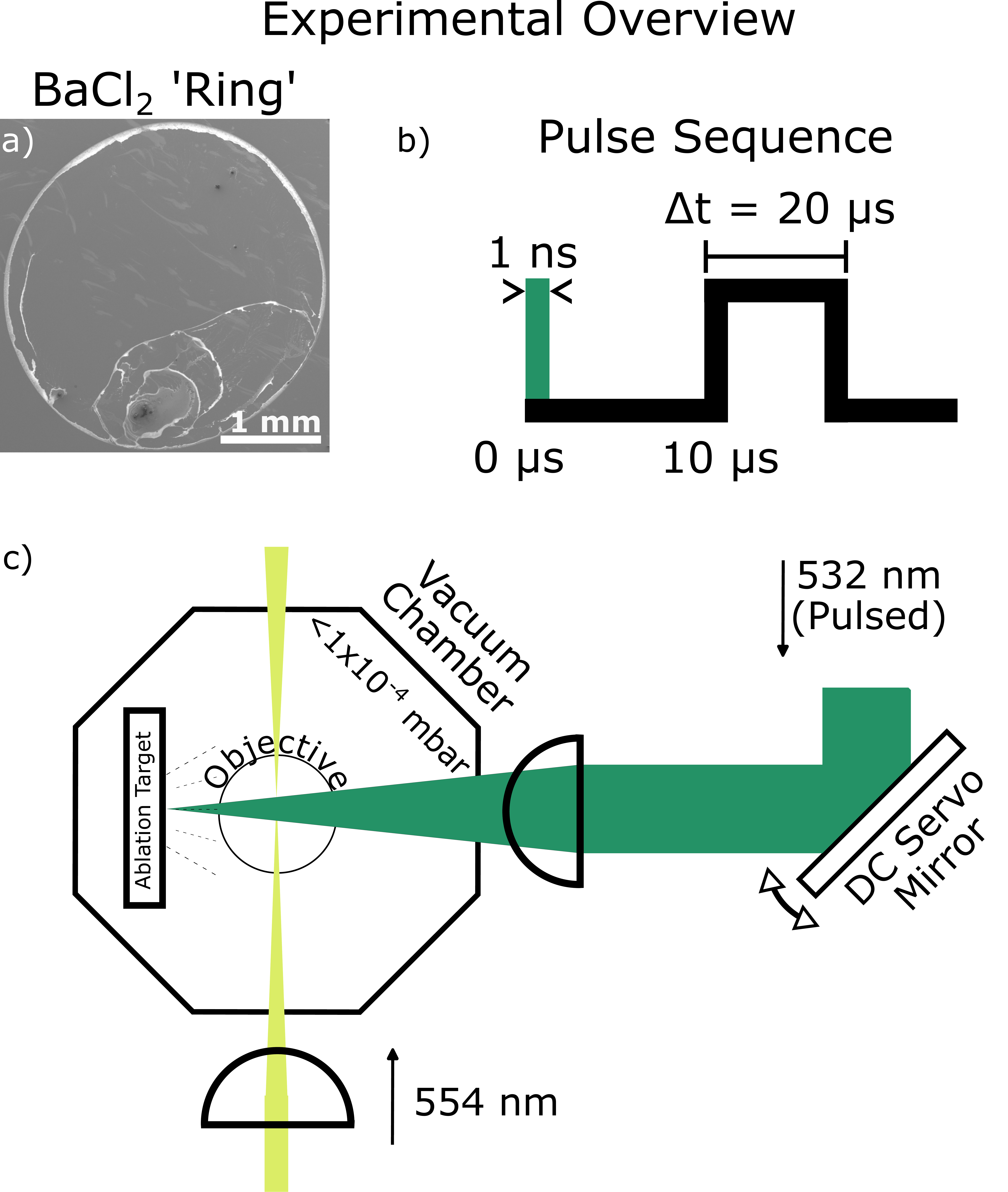}
\caption{\label{fig:fig1}
a) A ring of $\mathrm{BaCl}_2$ containing $\sim$~1-2 $\mu$g of material deposited on a gold microfabricated substrate. 
The microfabricated gold substrate is not an ablation target, but simply a smooth surface to estimate the amount of deposited $\mathrm{BaCl}_2$.
b) Pulse sequence of the experiment for collecting neutral fluorescence. 
The ablation pulse seen on the photo-multiplier tube (PMT) signals the start of the experiment, and the majority of the neutral atoms arrive roughly 10~$\mu$s later \cite{PhysRevA.105.033102}. 
The photon collection window is from 10-30~$\mu$s relative to the start of the experiment. 
c) Top-down view of the vacuum and optical system used to collect neutral fluorescence from the atomic plume after ablation of targets. 
An objective and PMT sit below the ablation testing chamber. 
}
\end{figure}

A major hurdle preventing consistent neutral barium production from microgram $\mathrm{BaCl}_2$ ablation targets are the concentrated $\mathrm{BaCl}_2$ `rings' that form during the deposition process, pictured in Fig. \ref{fig:fig1}a. 
The `coffee ring effect' is a well studied phenomena\cite{rey,das,HERTAEG202152} and we remedy this issue through the heat-treatment of microgram $\mathrm{BaCl}_2$ ablation targets.

We present two recipes for creating these targets, both of which use a form of heat-treatment to either melt the salt on the surface of the substrate or melt the underlying metallic substrate.
The first recipe resistively heats a strip of platinum to the melting point of $\mathrm{BaCl}_2$ ($\sim 960^{\circ}$ C), melting the salt on the surface \cite{Christensen}.
The second recipe is a novel process in which $\mathrm{BaCl}_2$ is combined with aluminum and heated in a vacuum chamber. 
In this manuscript, we test these two types of heat-treated targets, hereafter referred to as the Pt* and Al* ablation targets, and compare this with a Pt ablation target that has not been heat-treated. 
The ablation targets are pictured in the appendix.

After heat-treatment, we map the distribution of $\mathrm{BaCl}_2$ over the ablation target surfaces and yield of ablated $^{138}\mathrm{Ba}$ using techniques that are independent of the details of any specific ion trap, such as energy dispersive X-ray spectroscopy (EDS) and fluorescence-based detection.

We find that the uniformity of the ablation targets after heat-treatment is enhanced. 
Neutral fluorescence is detected from both non heat-treated and heat-treated microgram $\mathrm{BaCl}_2$ ablation targets, indicating their viability in trapped ion experiments. However, the consistency at which $^{138}\mathrm{Ba}$ atoms can be detected from the heat-treated targets exceeds that of ablation targets that have not undergone heat-treatment.
We show that heat-treated ablation targets are at least four times as likely to produce detectable amount of neutral $^{138}\mathrm{Ba}$ atoms and that neutral production can be sustained for at least three times as many ablation pulses while still producing measurable amounts of neutral fluorescence. 



\section{\label{sec:deposition} Deposition of $\mathrm{BaCl}_2$}
The first step in fabricating any microgram quantity $\mathrm{BaCl}_2$ ablation target is to deposit the salt onto a substrate. 
Radioactive $^{133}\mathrm{Ba}$ salt can be purchased from suppliers such as the Eckert $\&$ Ziegler Group and comes dissociated in a vial of solvent.
A buyer can choose the given amount of radioactivity in a vial, but for calculation purposes we assume the vial has $\sim 37$ MBq of radioactivity in order to avoid being near the CNSC $\sim 55$ MBq radiation limit.
Since the deposition presented here is meant to be compatible with the deposition of radioactive $\mathrm{BaCl}_2$, we create similar solutions with naturally abundant $\mathrm{BaCl}_2$ from Sigma-Aldrich (342920-10G) and deionized water. 
The procedure used to create these ablation targets is the same procedure is used for both the Pt* and Pt filament-like targets as well as the Al* targets.

The amount of $^{133}\mathrm{Ba}$ salt in a 37 MBq solution can be calculated from the half-life $t_{1/2}$ of the radioactive substance, the activity $A$, the reaction rate constant $\lambda$, and Avogadro's number $N_a$\cite{Nuclear}:   

\begin{eqnarray}
t_{1/2} = \frac{\ln (2)}{\lambda},
\\
A = \lambda N_a n 
\label{eq:one}.
\end{eqnarray}

The value of the known parameters $t_{1/2} \approx 3.33\cross~10^8$~s, $A = 3.7\cross 10^7$ Bq, and $N_a \approx 6.022\cross10^{23}$~mol\textsuperscript{-1} can be used to find the number of moles $n$ of $^{133}\mathrm{Ba}$. 
Given the molar mass of radioactive $\mathrm{BaCl}_2$ is $\sim~203.8$~g/mol\cite{ba133mass}, we calculate that there is $\sim~6$~$\mu$g of $\mathrm{BaCl}_2$ in a radioactive vial with the given parameters. 
We test targets with $\sim 5$ $\mu$g of nonradioactive $\mathrm{BaCl}_2$. 

\begin{figure*}
\includegraphics[width = 1\linewidth]{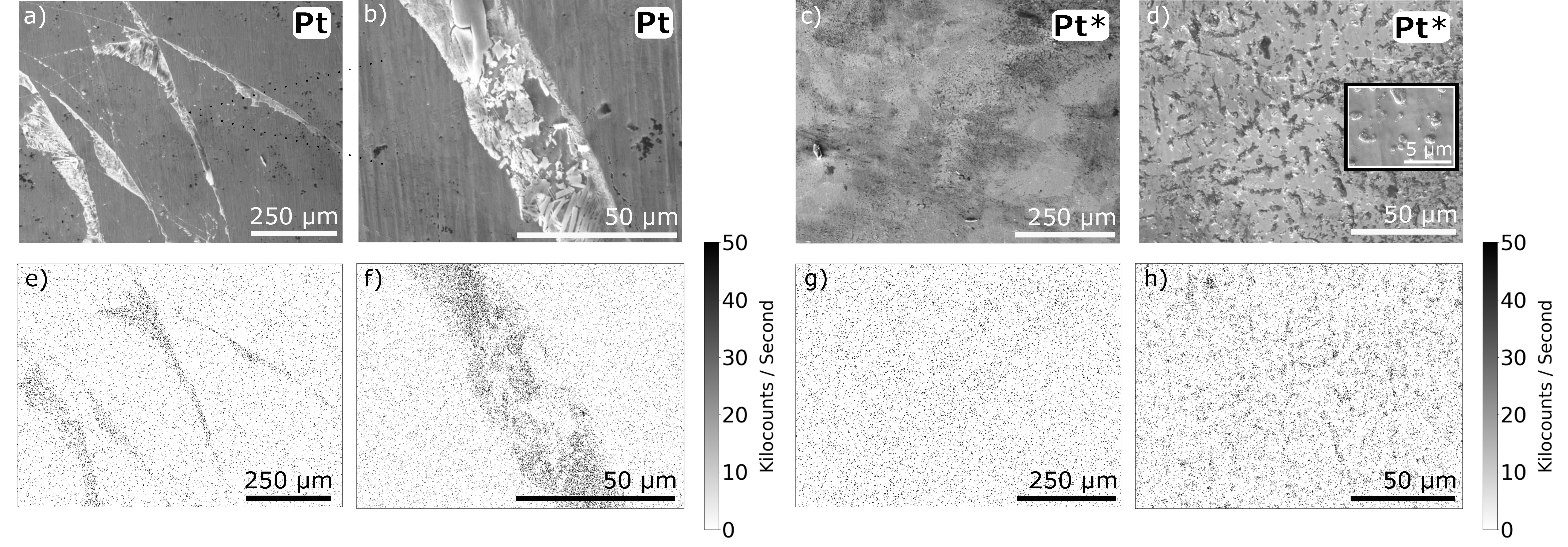}
\caption{\label{fig:fig2} a-d) SEM images of $\mathrm{BaCl}_2$ deposited onto platinum substrates near the center of the ablation targets. 
The $\mathrm{BaCl}_2$ of the Pt targets are concentrated into relatively thick rings, while the heat-treated Pt* ablation targets have $\mathrm{BaCl}_2$ uniformly spread over the entire surface in micron size crystals. 
e-h) Corresponding energy dispersive x-ray spectroscopy (EDS) maps of the above SEM images verifying the presence of barium after the deposition of $\mathrm{BaCl}_2$ and heat-treatment of the Pt* ablation targets.
All counts measured using EDS correspond to the barium $L_{\alpha1}$ line centered at 4.46 KeV. 
The Al* ablation targets are not mapped with EDS as the surface is quite rough and the underlying substrate (below the aluminum layer) is titanium with a $K_{\alpha1}$ line at 4.51 KeV, which is difficult to distinguish from barium. 
}
\end{figure*}

\subsection{\label{sec:depprep} Deposition Preparation $\&$ Process}

The substrate, whether it is a Pt*, Pt, or an Al* target is cleaned following typical ultra-high vacuum protocol.
The platinum is purchased from Sigma-Aldrich (327433) and the targets are cut into strips roughly 20 x 2.5 x 0.125 mm. 
The aluminum targets are made with 5-10 mg of aluminum foil from Kurt J. Lesker (FOILA24) melted onto a machined titanium holder.
We sonicate the pieces in Sparkleen™, rinse with deionized water, and sonicate again with acetone.

A 1-liter solution of deionized water is prepared with 20 mg of $\mathrm{BaCl}_2$.
The $\mathrm{BaCl}_2$ is weighed on a digital scale with 1 mg precision and added to the beaker of deionized water.
This is stirred until no more visible salt crystals are present and it has completely dissociated, usually within five minutes.  

We employ a 10-100 $\mu$L PuroPET variable volume micropipette with 10 $\mu$L disposable pipette tips to deposit 10 $\mu$L of solution at a time. 
We deposit 25 drops to ensure that $\sim$ 5 $\mu$g of salt is deposited.
One drop of solution is applied to the substrate at a time until the desired amount of $\mathrm{BaCl}_2$ has been applied. 

A hot plate is then set to $115^{\circ}$ C to increase the evaporation rate of the solvent.
At this temperature, we find 10 $\mu$L drops will not boil but evaporate at an increased rate.
The evaporation of a 10 $\mu$L drop takes about a minute. 
If the temperature set point is higher, the solution begins to boil vigorously, splattering $\mathrm{BaCl}_2$ and causing contamination and material loss. 

\section{\label{sec:heattreatment} Heat-treatment of $\mathrm{BaCl}_2$ Targets}

\subsection{\label{sec:resistive} Resistive Heating of Pt*}

After the deposition is complete, the ablation target is attached with alligator clips to a Chroma 62012P-80-60 programmable direct-current source and resistively heated by running a large amount of current $I\approx$~55~A through the target until the center of the filament glows bright orange and reaches temperatures $> 960^{\circ}$ C. 
This process is typically sustained for 10-20 seconds. 

\subsection{\label{sec:stage} High Vacuum Heating of Al*}

The second method of heat-treatment utilizes a high-vacuum BORALECTRIC® heating stage from Tectra GmbH, capable of reaching $1200^{\circ}$ C. 
The chamber is evacuated with a turbo-molecular pump until the pressure reaches $\leq$\num{1e-4} mbar to prevent oxidation of the target and heating stage.

The heating stage is ramped to the desired temperature of $760^{\circ}$ C.  
At this temperature a few milligrams of aluminum foil (MP = $660^{\circ}$ C) will melt within a couple of minutes once the machined target has come to equilibrium with the heating stage.  
The sample is visible through a viewport and the heat shield, so the phase transition can be visually confirmed. 
After the aluminum has melted, the stage is switched off and allowed to return $< 100^{\circ}$~C before venting and retrieving the samples. 
Once this first heat-treatment has occurred, $\mathrm{BaCl}_2$ is deposited on the surface of the Al target and the aluminum along with the $\mathrm{BaCl}_2$ is again heated to $\sim 760^{\circ}$ C in vacuum in order to diffuse the salt.
The aluminum is initially melted without $\mathrm{BaCl}_2$ in order to create a mechanically stable surface for depositing the salt. 

\section{\label{sec:measurementtech} Measurement Techniques}

\subsection{\label{sec:eds} EDS Measurements of $\mathrm{BaCl}_2$ Targets}
Commonly used in conjunction with a scanning electron microscope, EDS is a technique for probing the elemental composition on a sample surface\cite{edss}.
Using EDS after sample fabrication allows us to confirm the presence of $\mathrm{BaCl}_2$ on the surface of the Pt and Pt* targets and quantify the uniformity before ablation as shown in Fig.~\ref{fig:fig2}. 
The Pt targets depicted in Fig.~\ref{fig:fig2}a-b and Fig.~\ref{fig:fig2}e-f show concentrated areas of $\mathrm{BaCl}_2$.  
While the heat-treated Pt* targets in Fig.~\ref{fig:fig2}c-d and Fig.~\ref{fig:fig2}g-h show that the surface distribution of $\mathrm{BaCl}_2$ is relatively uniform.
We also map areas (not depicted) that are 100-150 $\mu$m in length.
We find that the sampled spots on the Pt* targets typically have $\sim$1-3$\%$ barium by weight on the surface before laser ablation, with an average of 2.5$\%$ and a coefficient of variation of 0.58.
A total of 53 spots were sampled on the Pt* target.

\begin{figure*}[t!]
\includegraphics[width = 1\linewidth]{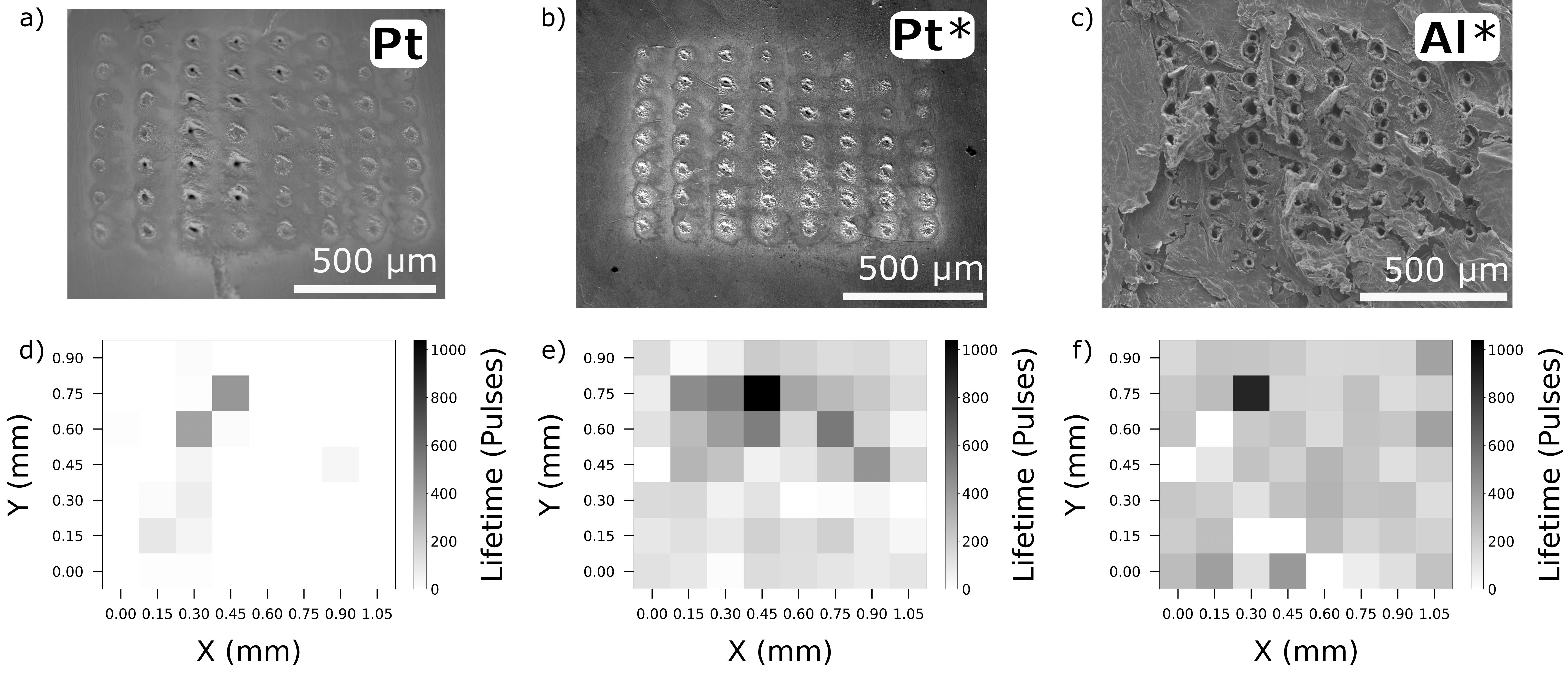}
\caption{\label{fig:fig3} a) Scanning electron microscope (SEM) picture of 56 ablation spots on a Pt ablation target that has not been heat-treated.
On this sample, 17 / 56 ablation spots produced measurable neutral fluorescence meeting the $\sigma_1$ criteria outlined in Section \ref{sec:spect}.
This is the best performing Pt target, with the other three Pt targets producing significantly fewer neutral atoms, despite ablating similar $\mathrm{BaCl}_2$ rings.
b) SEM image of a heat-treated Pt* ablation target with 55 / 56 spots producing enough neutral atoms to meet the $\sigma_1$ criteria.
c) SEM image of a heat-treated Al* ablation target with 51 / 56 spots producing enough neutral atoms to meet the $\sigma_1$ criteria.
d-f) Heatmaps of the ablation spot lifetime (number of pulses) corresponding to the above SEM images. 
Each ablated spot is spaced by $\sim 150$ $\mu$m, which is roughly the beam waist of the laser focal point on the ablation target.}
\end{figure*}

\begin{figure*}
\includegraphics[width = 1\linewidth]{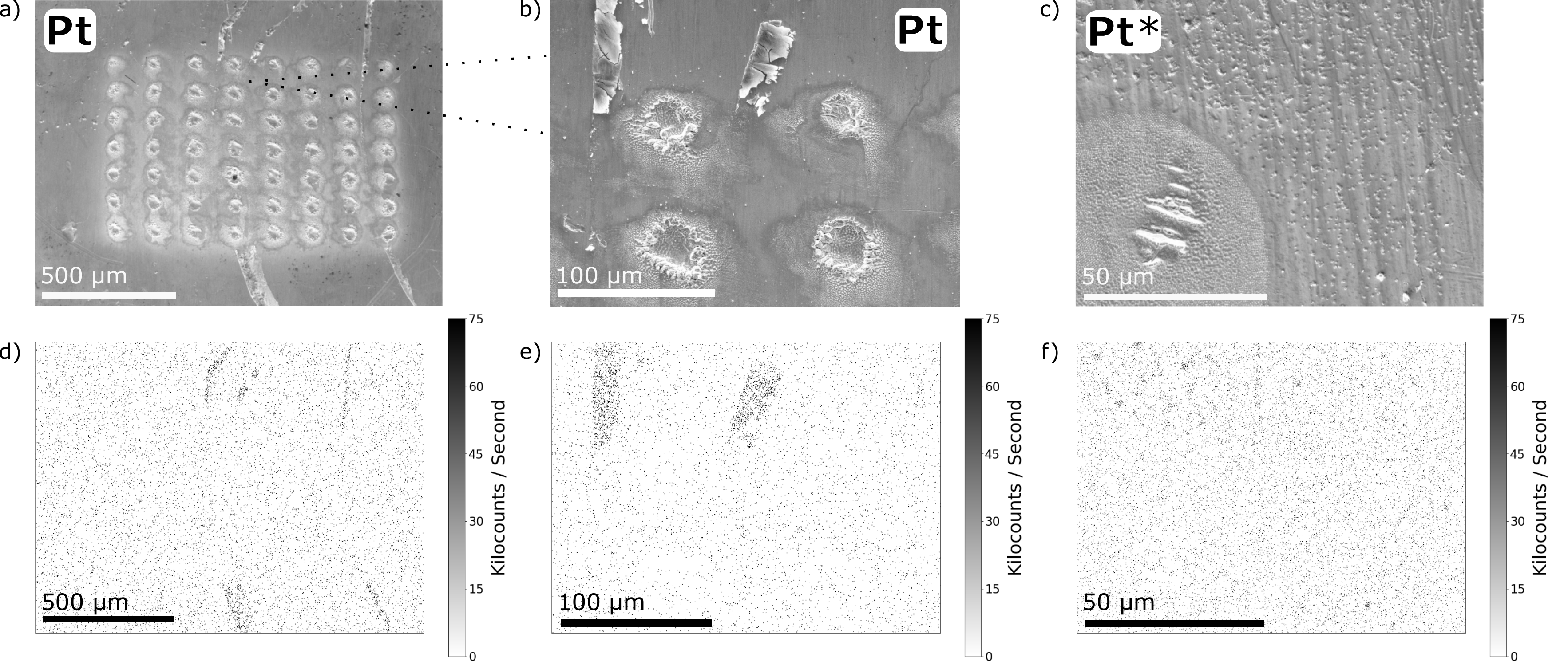}
\caption{\label{fig:fig4} a) A grid of 56 ablation spots covering $\sim 1.26$ mm\textsuperscript{2} on a Pt ablation target. Only $\sim$ 1 / 12 ablation spots aligned directly onto the $\mathrm{BaCl}_2$ ring satisfied the $\sigma_1$ criteria.
b) SEM image of $\mathrm{BaCl}_2$ rings that have been removed from the surface due to ablation without resulting in detection of neutral atoms.
c) SEM image of an ablated spot on a Pt* target. 
An ablation spot on the left side of the image has removed the $\sim 1$ $\mu$m $\mathrm{BaCl}_2$ crystals, but $\mathrm{BaCl}_2$ is still evident outside of the ablation spot.
The surface of the ablated spot is rough and appears similar to $\mathrm{BaCl}_2$ crystals, but it is devoid of any material except the platinum substrate. 
d-f) Corresponding EDS mappings of the ablated regions in SEM images shown in a-c), confirming the removal of the $\mathrm{BaCl}_2$ after ablation.
}
\end{figure*}

\subsection{\label{sec:fluor} Collection of Neutral Fluorescence from $^{138}\mathrm{Ba}$ Atoms}
We evaluate the Pt, Pt*, and Al* targets by collecting neutral fluorescence after ablation on a photo-multiplier tube (PMT), which can be correlated to the probability of loading\cite{PhysRevA.105.033102}. 
The pulse sequence used for collecting fluorescence is depicted in Fig.~\ref{fig:fig1}b. 
The ablation target is mounted 12 mm from the center of the chamber. 
There are two beams entering the chamber, one to drive the resonant $\ce{6s^2\:{}^1S_0} \leftrightarrow \ce{6s 6p\:{}^1P_1}$ transition at 554 nm and a perpendicular 532 nm Nd:YAG nanosecond pulsed solid-state laser from RPMC Lasers Inc. (WEDGE-HB 532) for ablation of the $\mathrm{BaCl}_2$. 
The 554 nm beam is aligned orthogonal to the normal of surface of the target to reduce Doppler broadening. 
After each ablation pulse ($\sim 0.42 \pm 0.02$ J/cm\textsuperscript{2} pulse fluence), counts are detected on a PMT from Hamamatsu (H10682-210) through a homebuilt 1" diameter objective (NA = 0.26, WD = 43 mm). 
The beam orientation and experimental setup is depicted in Fig.~\ref{fig:fig1}c. 
We consider only single-shot experiments, where only one ablation pulse is used to generate a cloud of neutral atoms. 
The ablation vacuum chamber from Kimball Physics (53-180000) is identical to the part used in previous ablation experiments involving a four-rod trap, as is the aforementioned objective\cite{PhysRevA.105.033102}. 

In order to discern counts from \textsuperscript{138}Ba atoms and counts from background events, we set the resonant frequency for driving the $\ce{6s^2\:{}^1S_0} \leftrightarrow \ce{6s 6p\:{}^1P_1}$ transition to 541.433000~THz and tune the 554 nm laser 1 GHz off-resonance to 541.432000~THz every other pulse.
The frequency of the 554 nm laser is stabilized via a HighFinesse WS8-2 wavemeter.
Neutral fluorescence is collected 10 $\mu$s after ablating the center of the target and the collection window lasts for 20 $\mu$s.

\section{\label{sec:spect} Results}
We ablate targets prepared with 5 $\mu$g of $\mathrm{BaCl}_2$ and search for neutral fluorescence produced from the ablation targets as a means to gauge the viability of the targets for use in trapped ion experiments. 
Both consistency and longevity of spots on the ablation targets are compared across the three target types as shown in Fig.~\ref{fig:fig3}.
The average spot lifetime of the Pt target shown in Fig.~\ref{fig:fig3}d is $\sim$ 69 pulses, compared with $\sim$ 185 and $\sim$ 232 pulses for the Pt* and Al* targets depicted in Fig.~\ref{fig:fig3}e and Fig.~\ref{fig:fig3}f, respectively.

The lifetime of an ablation spot is evaluated based on a running average of the counts collected shortly after the last 50 ablation pulses.
In this case, 25 pulses are when the 554 nm laser frequency is on-resonance and 25 pulses are when the frequency is tuned off-resonance.
All ablation spots are evaluated for a minimum of 50 pulses.
If the average of the counts on-resonance $c_R$ falls within one standard deviation $\sigma_1$ or two standard deviations $\sigma_2$ of the off-resonance $c_O$ count average, the spot is considered `dead' and unlikely to produce detectable \textsuperscript{138}Ba atoms. 
The lifetime of an ablation spot can be evaluated using either the $\sigma_1$ or $\sigma_2$ criteria.
A spot is considered `dead' spot if it satisfies Eq.~\ref{dead}.

\begin{equation}
\label{dead}
    c_R \leq c_O + \sigma_n 
\end{equation}
Over the course of an experiment, the standard deviation of the 554 nm laser frequency is $<$ 2 MHz shot-to-shot. 
The saturation parameter is set to $s > 100$ for the $\ce{6s^2\:{}^1S_0}~\leftrightarrow~\ce{6s 6p\:{}^1P_1}$ transition when collecting neutral fluorescence to ensure sufficient power.

\begin{table}[H]
\begin{center}
    \begin{tabular}{|c|c|c|c|c|c|}
    \toprule\toprule    
    \hline
    Target& $c_1$ (Pulses) & Yield ($\sigma_1$ Criteria) & $c_2$ (Pulses) & Yield ($\sigma_2$ Criteria)\\ \hline
    Pt & 36 & 43 / 79\textsuperscript{\textdagger} $\approx$ 0.54 & 32 & 10 / 79 $\approx$ 0.13 \\ 
    Pt* & 158 & 142 / 168 $\approx$ 0.85 &  91 & 95 / 168 $\approx$ 0.57 \\ 
    Al* & 184 & 134 / 175 $\approx$ 0.77 & 123 & 93 / 175 $\approx$ 0.53 \\  \hline\bottomrule\bottomrule
    \end{tabular}
    \begin{flushleft}
  
    \textdagger: Only 79 / 448 total spots on the Pt type targets hit a $\mathrm{BaCl}_2$ ring, so only these ablation spots are considered in the average lifetime calculations.
    In realistic scenarios, it is not possible to deterministically hit the ring, so the practical yield is 43 / 448 $\approx$ 0.10.
    \end{flushleft}
    \caption{Mean lifetime $c_1$ of all ablation spots on each target that satisfies the $\sigma_1$ criteria in Section \ref{sec:spect}. 
    Given that the heat-treated Pt* and Al* ablation targets have a significantly more uniform distribution of $\mathrm{BaCl}_2$ on the surface, all spots on these target types were considered in the average lifetime calculations.
    Mean lifetime $c_2$ of only the spots that produced significant neutral fluorescence as defined by the $\sigma_2$ criteria.
    The fraction of tested ablation spots that meet the $\sigma_1$ and $\sigma_2$ criteria are also listed for each ablation target type.
    }
    \label{table:lifetimes}
\end{center}
\end{table}


A total of ten ablation targets were tested: three targets for each of the heat-treated Pt* and Al* type targets and four of the Pt type target, listed in Table \ref{table:lifetimes}.
A minimum of 56 spots were collected from each of the ten targets.
The heat-treated Pt* and Al* targets performed significantly better in both consistency and \textsuperscript{138}Ba production on average when compared with the Pt targets that have not undergone heat-treatment. 
We find that despite ablating directly on the $\mathrm{BaCl}_2$ `rings' of the Pt targets, often there is no discernible neutral fluorescence collected above the background. 
We confirm with SEM and EDS measurements that $\mathrm{BaCl}_2$ is removed from the surface of the Pt and Pt* targets at ablated spots, as depicted in Fig.~\ref{fig:fig4}, but only a single spot from the Pt target shown in Fig. 4 yielded measurable neutral counts. 

\section{\label{sec:conclusion}Conclusion} 
We have outlined and tested two heat-treatment techniques for fabricating microgram quantity $\mathrm{BaCl}_2$ ablation targets to be used in isotope-selective loading of barium ions, motivated by trapping efforts aimed at the $^{133}\mathrm{Ba}^+$ isotope. 
With both heat-treatment techniques, a more uniform distribution of the $\mathrm{BaCl}_2$ was created on the surface of the substrate, creating targets that are at least four times more likely to produce neutral atoms. 
The total yield of ablated barium atoms is also significantly higher with ablation spot lifetimes three to five times higher for heat-treated targets.
Similarly, when testing ablation targets with $> 100$ $\mu$g of $\mathrm{BaCl}_2$, the uniformity of the salt on the surface is improved after heat-treatment. 
More details on the uniformity of the $> 100$ $\mu$g targets, as well as observations on contamination of nearby surfaces by ablated $\mathrm{BaCl}_2$ can be found in the appendices. 

We find that in the case of the platinum substrate ablation targets, consistent production of neutral fluorescence is increased significantly when the substrate is heated to temperatures that enable melting of the microgram quantities of $\mathrm{BaCl}_2$.
This is due to the fact that, when melted, concentrated rings of $\mathrm{BaCl}_2$ that form during the deposition process are dispersed over the substrate surface. 
Given the spatial uniformity of the $\mathrm{BaCl}_2$ on the surface of the Pt* ablation targets and the ease of preparation compared with the Al* targets, we find that of the tested approaches the Pt* target performs with the best consistency.
Our presented recipe and results enable a more consistent preparation of $\mathrm{BaCl}_2$ ablation targets, paving the way for integration of the $^{133}\mathrm{Ba}^+$ isotope into scalable quantum information experiments. 
Due to the inherent advantages $^{133}\mathrm{Ba}^+$ offers for quantum information processing, our research removes a major obstacle toward advancing trapped ion quantum computing technology. 

\begin{acknowledgments}
This research was supported in part by the University of Waterloo, Canada First Research Excellence Fund (CFREF), Grant No. CFREF-2015-00011, and the Ontario Early Researcher Award. CS is also supported by a Canada Research Chair.

We thank Zachary J. Wall and Samuel R. Vizvary, as well as the other group members at the University of California-Los Angeles for stimulating discussion and transparency involving their successful loading of $^{133}\mathrm{Ba}^+$.
\end{acknowledgments}

\appendix

\section{Procurement of $^{133}\mathrm{Ba}$}

The $^{133}\mathrm{Ba}$ isotope can be purchased from Eckert $\&$ Ziegler, Oak Ridge National Lab, or NIST. 
For future experiments, we have purchased $^{133}\mathrm{Ba}$ from Eckert $\&$ Ziegler with "no stable carrier added" and request that the sample be pulled from the master batch with the highest "specific radioactivity".

\section{Aluminum Heat-Treatment $\&$ Ablation Targets}
\label{appb}
We present a figure showing targets that were ablated during the experiment as well as a rendering of the high-vacuum chamber and heating stage that was used to manufacture the Al* ablation targets.  

\begin{figure}
\includegraphics[width =\linewidth]{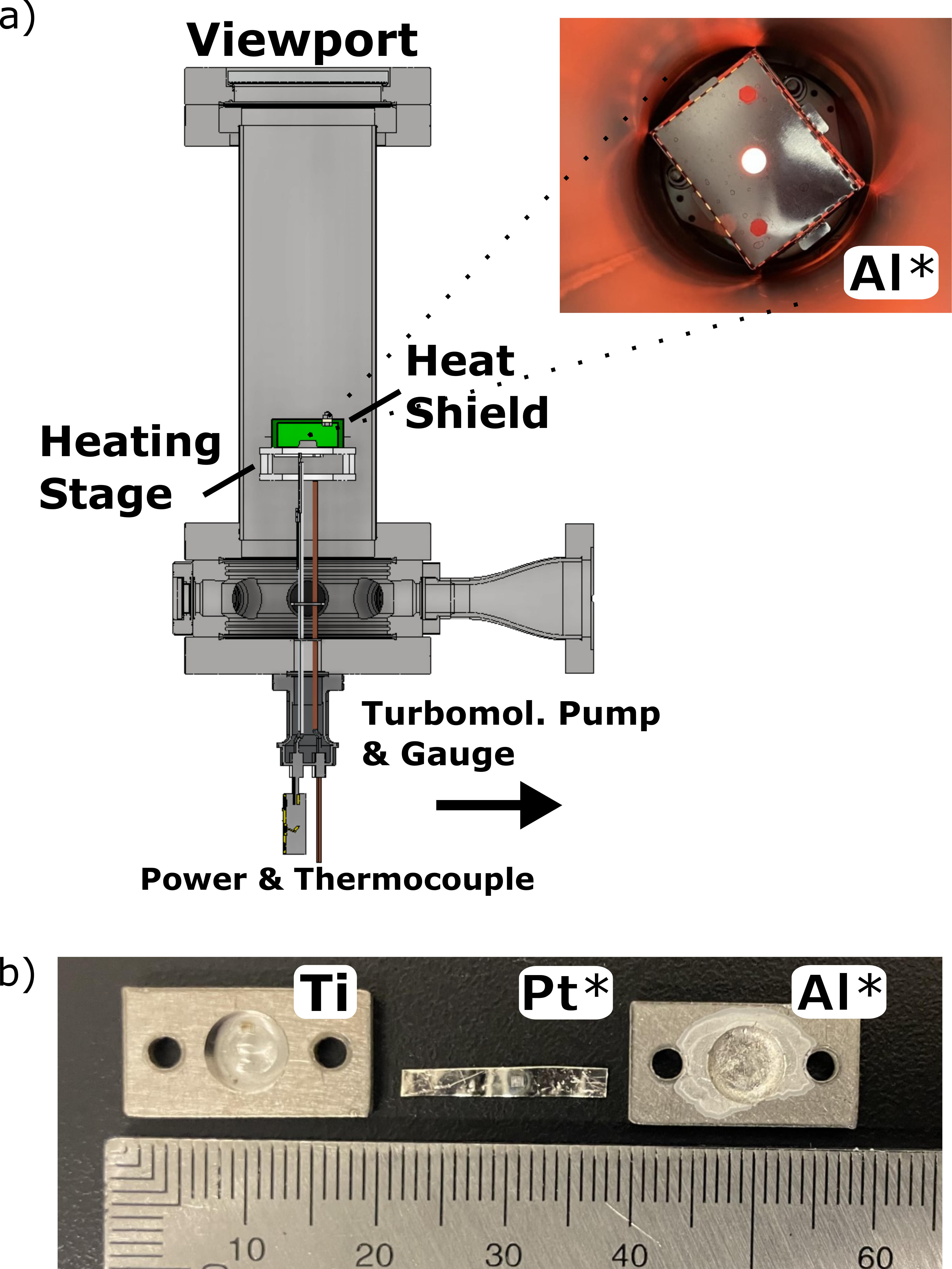}
\caption{\label{fig:fig5} a) Cross-sectional view of the high-vacuum heating stage used to create Al* ablation targets with microgram quantities of $\mathrm{BaCl}_2$.
b) Tested ablation targets with $\sim$ 5 $\mu$g of $\mathrm{BaCl}_2$ on the surface. 
The appearance of the heat-treated Pt* and Pt targets are similar.
The ruler depicted is in millimeters.}
\end{figure}

\section{Trap Contamination from $\mathrm{BaCl}_2$ Ablation Targets}

Ablation of a non heat-treated natural abundance $\mathrm{BaCl}_2$ targets was also conducted in an ultra-high vacuum chamber ($\sim 4.4\times10^{-11}$ mbar) where a surface trap \cite{revelle2020phoenix} was installed $\sim 15$ mm away from a macroscopic $\mathrm{BaCl}_2$ target containing $\sim 40$ mg worth of salt. 
The surface trap is a defected test trap with many electrodes shorted to ground. 
We collected neutral fluorescence from $^{138}\mathrm{Ba}$ as we ablated the target.
After confirming the presence of $^{138}\mathrm{Ba}$ at the trapping region, the physical test trap was removed and imaged using an SEM. 

\begin{figure*}
\includegraphics[width = 0.85\linewidth]{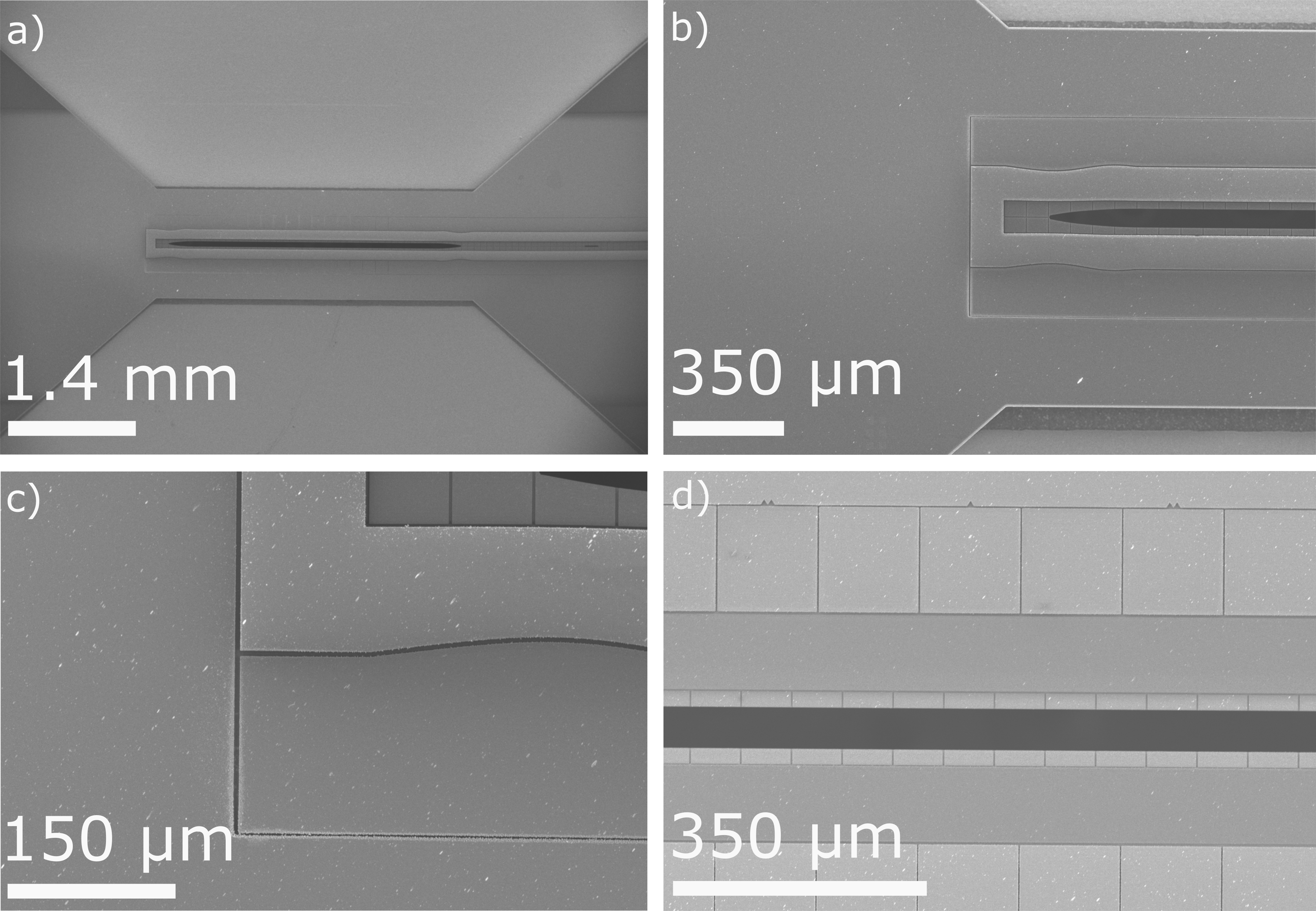}
\caption{\label{fig:fig6} a-d) SEM images of the test trap with large white $\mathrm{BaCl}_2$ crystals visible from afar, as well as smaller crystals on outer and inner DC electrodes. 
The $\mathrm{BaCl}_2$ crystal contamination is directional, but the shear number of crystals on the surface is evident. 
The individual $\mathrm{BaCl}_2$ crystal sizes usually range from $1-10$ $\mu$m in length.
We estimate this contamination is after $>$ 50,000 ablation pulses using a similar ablation fluence to that as stated in the manuscript.}
\end{figure*}

Significant trap contamination was found, visible on SEM images after ablation of macroscopic $\mathrm{BaCl}_2$ targets, as shown in Fig. \ref{fig:fig6}.
$\mathrm{BaCl}_2$ is a dielectric, so the chances of it directly shorting electrodes is unlikely.
However, excess material deposited on the surface of microfabricated traps can exacerbate anomalous heating of the trapped ion motion.
This is an area of considerable focus \cite{PhysRevA.61.063418, PhysRevLett.97.103007} and a limiting factor in making even smaller micro-fabricated surface traps \cite{PhysRevA.97.020302}.

\section{Heat-Treated Macroscopic Targets}

\begin{figure}
\includegraphics[width = \linewidth]{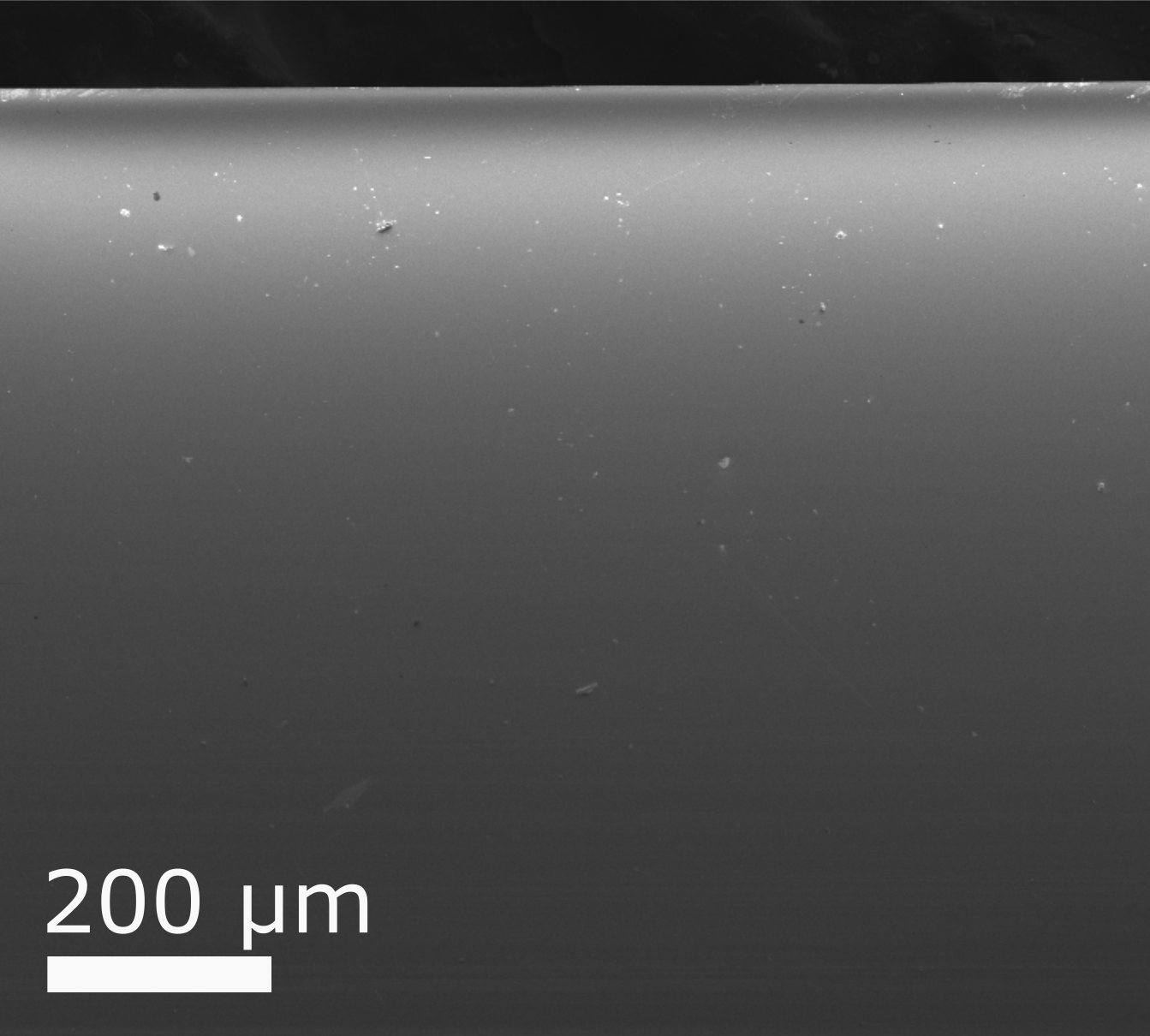}
\caption{\label{fig:fig7} SEM image of a gold prefabricated chip with $\mathrm{BaCl}_2$ collected on the surface after ablation of macroscopic targets. 
This is from a $\sim 200$ $\mu$g $\mathrm{BaCl}_2$ ablation target, proving that even after heat-treatment that small crystals can be ejected from the target without sublimation.}
\end{figure}

\begin{figure*}
\includegraphics[width = \linewidth]{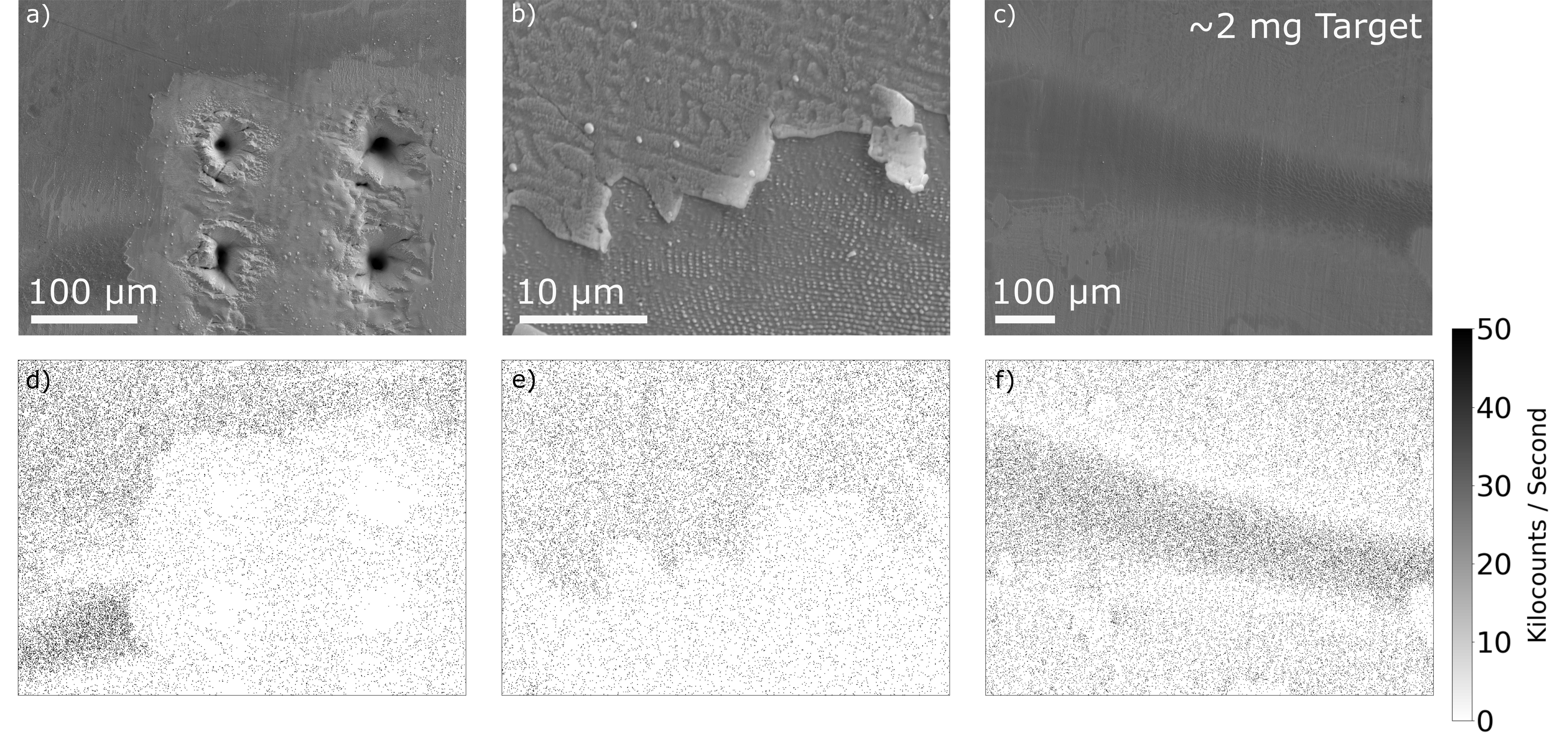}
\caption{\label{fig:fig8} a-c) SEM images of the heat-treated platinum ablation targets with $\sim 2$ mg of $\mathrm{BaCl}_2$ deposited on the substrate. 
Note that a layer of $\mathrm{BaCl}_2$ is formed as a crust on the surface of the platinum, making the ablation of these targets more consistent.
The crust is removed after ablation.
The size of the platinum strip is similar to the strip in Fig.~\ref{fig:fig5}b.}
\end{figure*}

We test heat-treated platinum targets with macroscopic (visible) varying amounts of $\mathrm{BaCl}_2$ in order to evaluate if heat-treatment can eliminate contamination of surface traps when using $\mathrm{BaCl}_2$ ablation targets.
We present scanning electron microscope (SEM) and energy dispersive x-ray spectroscopy (EDS) images of the targets after ablation in Fig.~\ref{fig:fig8}.
Neutral fluorescence is collected from all ablation targets in the same manner described in the manuscript, with a total of 48 spots being ablated (16 for each target). 
We find that melting $> 100$ $\mu$g of $\mathrm{BaCl}_2$ on platinum substrates produces uniform ablation targets with spots that last $> 1000$ pulses, but that contamination of a surface trap would likely still occur as evident in Fig.~\ref{fig:fig7}. 

We ablate heat-treated platinum targets with $\sim$ 200 $\mu$g, 2 mg, and $>$ 20 mg of $\mathrm{BaCl}_2$. 
The deposition and heat-treatment follows the recipe outlined in the manuscript for these specific target types.
The ablation targets are installed in the high-vacuum test chamber along with microfabricated gold chips from Angstrom Engineering Inc. (QA05-00180).
The chips are $\sim 25$ mm away from the ablation targets in order to re-create a similar situation with a microfabricated surface trap.
After ablation, the gold chips are removed and imaged as shown in Fig.~\ref{fig:fig7}.
 
\nocite{*}
\bibliography{aipsamp}

\begin{thebibliography}{28}%
\makeatletter
\providecommand \@ifxundefined [1]{%
 \@ifx{#1\undefined}
}%
\providecommand \@ifnum [1]{%
 \ifnum #1\expandafter \@firstoftwo
 \else \expandafter \@secondoftwo
 \fi
}%
\providecommand \@ifx [1]{%
 \ifx #1\expandafter \@firstoftwo
 \else \expandafter \@secondoftwo
 \fi
}%
\providecommand \natexlab [1]{#1}%
\providecommand \enquote  [1]{``#1''}%
\providecommand \bibnamefont  [1]{#1}%
\providecommand \bibfnamefont [1]{#1}%
\providecommand \citenamefont [1]{#1}%
\providecommand \href@noop [0]{\@secondoftwo}%
\providecommand \href [0]{\begingroup \@sanitize@url \@href}%
\providecommand \@href[1]{\@@startlink{#1}\@@href}%
\providecommand \@@href[1]{\endgroup#1\@@endlink}%
\providecommand \@sanitize@url [0]{\catcode `\\12\catcode `\$12\catcode `\&12\catcode `\#12\catcode `\^12\catcode `\_12\catcode `\%12\relax}%
\providecommand \@@startlink[1]{}%
\providecommand \@@endlink[0]{}%
\providecommand \url  [0]{\begingroup\@sanitize@url \@url }%
\providecommand \@url [1]{\endgroup\@href {#1}{\urlprefix }}%
\providecommand \urlprefix  [0]{URL }%
\providecommand \Eprint [0]{\href }%
\providecommand \doibase [0]{http://dx.doi.org/}%
\providecommand \selectlanguage [0]{\@gobble}%
\providecommand \bibinfo  [0]{\@secondoftwo}%
\providecommand \bibfield  [0]{\@secondoftwo}%
\providecommand \translation [1]{[#1]}%
\providecommand \BibitemOpen [0]{}%
\providecommand \bibitemStop [0]{}%
\providecommand \bibitemNoStop [0]{.\EOS\space}%
\providecommand \EOS [0]{\spacefactor3000\relax}%
\providecommand \BibitemShut  [1]{\csname bibitem#1\endcsname}%
\let\auto@bib@innerbib\@empty
\bibitem [{\citenamefont {Monroe}\ \emph {et~al.}(2014)\citenamefont {Monroe}, \citenamefont {Raussendorf}, \citenamefont {Ruthven}, \citenamefont {Brown}, \citenamefont {Maunz}, \citenamefont {Duan},\ and\ \citenamefont {Kim}}]{PhysRevA.89.022317}%
  \BibitemOpen
  \bibfield  {author} {\bibinfo {author} {\bibfnamefont {C.}~\bibnamefont {Monroe}}, \bibinfo {author} {\bibfnamefont {R.}~\bibnamefont {Raussendorf}}, \bibinfo {author} {\bibfnamefont {A.}~\bibnamefont {Ruthven}}, \bibinfo {author} {\bibfnamefont {K.~R.}\ \bibnamefont {Brown}}, \bibinfo {author} {\bibfnamefont {P.}~\bibnamefont {Maunz}}, \bibinfo {author} {\bibfnamefont {L.-M.}\ \bibnamefont {Duan}}, \ and\ \bibinfo {author} {\bibfnamefont {J.}~\bibnamefont {Kim}},\ }\href {\doibase 10.1103/PhysRevA.89.022317} {\bibfield  {journal} {\bibinfo  {journal} {Phys. Rev. A}\ }\textbf {\bibinfo {volume} {89}},\ \bibinfo {pages} {022317} (\bibinfo {year} {2014})}\BibitemShut {NoStop}%
\bibitem [{\citenamefont {Kielpinski}\ \emph {et~al.}(2002)\citenamefont {Kielpinski}, \citenamefont {Monroe},\ and\ \citenamefont {Wineland}}]{article1}%
  \BibitemOpen
  \bibfield  {author} {\bibinfo {author} {\bibfnamefont {D.}~\bibnamefont {Kielpinski}}, \bibinfo {author} {\bibfnamefont {C.}~\bibnamefont {Monroe}}, \ and\ \bibinfo {author} {\bibfnamefont {D.}~\bibnamefont {Wineland}},\ }\href {\doibase 10.1038/nature00784} {\bibfield  {journal} {\bibinfo  {journal} {Nature}\ }\textbf {\bibinfo {volume} {417}},\ \bibinfo {pages} {709} (\bibinfo {year} {2002})}\BibitemShut {NoStop}%
\bibitem [{\citenamefont {Ballance}\ \emph {et~al.}(2016)\citenamefont {Ballance}, \citenamefont {Harty}, \citenamefont {Linke}, \citenamefont {Sepiol},\ and\ \citenamefont {Lucas}}]{PhysRevLett.117.060504}%
  \BibitemOpen
  \bibfield  {author} {\bibinfo {author} {\bibfnamefont {C.~J.}\ \bibnamefont {Ballance}}, \bibinfo {author} {\bibfnamefont {T.~P.}\ \bibnamefont {Harty}}, \bibinfo {author} {\bibfnamefont {N.~M.}\ \bibnamefont {Linke}}, \bibinfo {author} {\bibfnamefont {M.~A.}\ \bibnamefont {Sepiol}}, \ and\ \bibinfo {author} {\bibfnamefont {D.~M.}\ \bibnamefont {Lucas}},\ }\href {\doibase 10.1103/PhysRevLett.117.060504} {\bibfield  {journal} {\bibinfo  {journal} {Phys. Rev. Lett.}\ }\textbf {\bibinfo {volume} {117}},\ \bibinfo {pages} {060504} (\bibinfo {year} {2016})}\BibitemShut {NoStop}%
\bibitem [{\citenamefont {Boguslawski}\ \emph {et~al.}(2023)\citenamefont {Boguslawski}, \citenamefont {Wall}, \citenamefont {Vizvary}, \citenamefont {Moore}, \citenamefont {Bareian}, \citenamefont {Allcock}, \citenamefont {Wineland}, \citenamefont {Hudson},\ and\ \citenamefont {Campbell}}]{PhysRevLett.131.063001}%
  \BibitemOpen
  \bibfield  {author} {\bibinfo {author} {\bibfnamefont {M.~J.}\ \bibnamefont {Boguslawski}}, \bibinfo {author} {\bibfnamefont {Z.~J.}\ \bibnamefont {Wall}}, \bibinfo {author} {\bibfnamefont {S.~R.}\ \bibnamefont {Vizvary}}, \bibinfo {author} {\bibfnamefont {I.~D.}\ \bibnamefont {Moore}}, \bibinfo {author} {\bibfnamefont {M.}~\bibnamefont {Bareian}}, \bibinfo {author} {\bibfnamefont {D.~T.~C.}\ \bibnamefont {Allcock}}, \bibinfo {author} {\bibfnamefont {D.~J.}\ \bibnamefont {Wineland}}, \bibinfo {author} {\bibfnamefont {E.~R.}\ \bibnamefont {Hudson}}, \ and\ \bibinfo {author} {\bibfnamefont {W.~C.}\ \bibnamefont {Campbell}},\ }\href {\doibase 10.1103/PhysRevLett.131.063001} {\bibfield  {journal} {\bibinfo  {journal} {Phys. Rev. Lett.}\ }\textbf {\bibinfo {volume} {131}},\ \bibinfo {pages} {063001} (\bibinfo {year} {2023})}\BibitemShut {NoStop}%
\bibitem [{\citenamefont {Wang}\ \emph {et~al.}(2021)\citenamefont {Wang}, \citenamefont {Luan}, \citenamefont {Qiao}, \citenamefont {Um}, \citenamefont {Zhang}, \citenamefont {Wang}, \citenamefont {Yuan}, \citenamefont {Gu}, \citenamefont {Zhang},\ and\ \citenamefont {Kim}}]{Wang2021-es}%
  \BibitemOpen
  \bibfield  {author} {\bibinfo {author} {\bibfnamefont {P.}~\bibnamefont {Wang}}, \bibinfo {author} {\bibfnamefont {C.-Y.}\ \bibnamefont {Luan}}, \bibinfo {author} {\bibfnamefont {M.}~\bibnamefont {Qiao}}, \bibinfo {author} {\bibfnamefont {M.}~\bibnamefont {Um}}, \bibinfo {author} {\bibfnamefont {J.}~\bibnamefont {Zhang}}, \bibinfo {author} {\bibfnamefont {Y.}~\bibnamefont {Wang}}, \bibinfo {author} {\bibfnamefont {X.}~\bibnamefont {Yuan}}, \bibinfo {author} {\bibfnamefont {M.}~\bibnamefont {Gu}}, \bibinfo {author} {\bibfnamefont {J.}~\bibnamefont {Zhang}}, \ and\ \bibinfo {author} {\bibfnamefont {K.}~\bibnamefont {Kim}},\ }\href@noop {} {\bibfield  {journal} {\bibinfo  {journal} {Nature Communications}\ }\textbf {\bibinfo {volume} {12}},\ \bibinfo {pages} {233} (\bibinfo {year} {2021})}\BibitemShut {NoStop}%
\bibitem [{\citenamefont {An}\ \emph {et~al.}(2022)\citenamefont {An}, \citenamefont {Ransford}, \citenamefont {Schaffer}, \citenamefont {Sletten}, \citenamefont {Gaebler}, \citenamefont {Hostetter},\ and\ \citenamefont {Vittorini}}]{PhysRevLett.129.130501}%
  \BibitemOpen
  \bibfield  {author} {\bibinfo {author} {\bibfnamefont {F.~A.}\ \bibnamefont {An}}, \bibinfo {author} {\bibfnamefont {A.}~\bibnamefont {Ransford}}, \bibinfo {author} {\bibfnamefont {A.}~\bibnamefont {Schaffer}}, \bibinfo {author} {\bibfnamefont {L.~R.}\ \bibnamefont {Sletten}}, \bibinfo {author} {\bibfnamefont {J.}~\bibnamefont {Gaebler}}, \bibinfo {author} {\bibfnamefont {J.}~\bibnamefont {Hostetter}}, \ and\ \bibinfo {author} {\bibfnamefont {G.}~\bibnamefont {Vittorini}},\ }\href {\doibase 10.1103/PhysRevLett.129.130501} {\bibfield  {journal} {\bibinfo  {journal} {Phys. Rev. Lett.}\ }\textbf {\bibinfo {volume} {129}},\ \bibinfo {pages} {130501} (\bibinfo {year} {2022})}\BibitemShut {NoStop}%
\bibitem [{\citenamefont {Christensen}\ \emph {et~al.}(2020)\citenamefont {Christensen}, \citenamefont {Hucul}, \citenamefont {Campbell},\ and\ \citenamefont {Hudson}}]{Christensen2020}%
  \BibitemOpen
  \bibfield  {author} {\bibinfo {author} {\bibfnamefont {J.~E.}\ \bibnamefont {Christensen}}, \bibinfo {author} {\bibfnamefont {D.}~\bibnamefont {Hucul}}, \bibinfo {author} {\bibfnamefont {W.~C.}\ \bibnamefont {Campbell}}, \ and\ \bibinfo {author} {\bibfnamefont {E.~R.}\ \bibnamefont {Hudson}},\ }\href {\doibase 10.1038/s41534-020-0265-5} {\bibfield  {journal} {\bibinfo  {journal} {npj Quantum Information}\ }\textbf {\bibinfo {volume} {6}},\ \bibinfo {pages} {35} (\bibinfo {year} {2020})}\BibitemShut {NoStop}%
\bibitem [{\citenamefont {Hucul}\ \emph {et~al.}(2017)\citenamefont {Hucul}, \citenamefont {Christensen}, \citenamefont {Hudson},\ and\ \citenamefont {Campbell}}]{PhysRevLett.119.100501}%
  \BibitemOpen
  \bibfield  {author} {\bibinfo {author} {\bibfnamefont {D.}~\bibnamefont {Hucul}}, \bibinfo {author} {\bibfnamefont {J.~E.}\ \bibnamefont {Christensen}}, \bibinfo {author} {\bibfnamefont {E.~R.}\ \bibnamefont {Hudson}}, \ and\ \bibinfo {author} {\bibfnamefont {W.~C.}\ \bibnamefont {Campbell}},\ }\href {\doibase 10.1103/PhysRevLett.119.100501} {\bibfield  {journal} {\bibinfo  {journal} {Phys. Rev. Lett.}\ }\textbf {\bibinfo {volume} {119}},\ \bibinfo {pages} {100501} (\bibinfo {year} {2017})}\BibitemShut {NoStop}%
\bibitem [{\citenamefont {Binai-Motlagh}\ \emph {et~al.}(2023)\citenamefont {Binai-Motlagh}, \citenamefont {Day}, \citenamefont {Videnov}, \citenamefont {Greenberg}, \citenamefont {Senko},\ and\ \citenamefont {Islam}}]{Binai_Motlagh_2023}%
  \BibitemOpen
  \bibfield  {author} {\bibinfo {author} {\bibfnamefont {A.}~\bibnamefont {Binai-Motlagh}}, \bibinfo {author} {\bibfnamefont {M.~L.}\ \bibnamefont {Day}}, \bibinfo {author} {\bibfnamefont {N.}~\bibnamefont {Videnov}}, \bibinfo {author} {\bibfnamefont {N.}~\bibnamefont {Greenberg}}, \bibinfo {author} {\bibfnamefont {C.}~\bibnamefont {Senko}}, \ and\ \bibinfo {author} {\bibfnamefont {R.}~\bibnamefont {Islam}},\ }\href {\doibase 10.1088/2058-9565/ace6cb} {\bibfield  {journal} {\bibinfo  {journal} {Quantum Science and Technology}\ }\textbf {\bibinfo {volume} {8}},\ \bibinfo {pages} {045012} (\bibinfo {year} {2023})}\BibitemShut {NoStop}%
\bibitem [{\citenamefont {Low}\ \emph {et~al.}(2023)\citenamefont {Low}, \citenamefont {White},\ and\ \citenamefont {Senko}}]{low2023control}%
  \BibitemOpen
  \bibfield  {author} {\bibinfo {author} {\bibfnamefont {P.~J.}\ \bibnamefont {Low}}, \bibinfo {author} {\bibfnamefont {B.}~\bibnamefont {White}}, \ and\ \bibinfo {author} {\bibfnamefont {C.}~\bibnamefont {Senko}},\ }\href@noop {} {\enquote {\bibinfo {title} {Control and readout of a 13-level trapped ion qudit},}\ } (\bibinfo {year} {2023}),\ \Eprint {http://arxiv.org/abs/2306.03340} {arXiv:2306.03340 [quant-ph]} \BibitemShut {NoStop}%
\bibitem [{\citenamefont {Vizvary}\ \emph {et~al.}(2023)\citenamefont {Vizvary}, \citenamefont {Wall}, \citenamefont {Boguslawski}, \citenamefont {Bareian}, \citenamefont {Derevianko}, \citenamefont {Campbell},\ and\ \citenamefont {Hudson}}]{vizvary2023eliminating}%
  \BibitemOpen
  \bibfield  {author} {\bibinfo {author} {\bibfnamefont {S.~R.}\ \bibnamefont {Vizvary}}, \bibinfo {author} {\bibfnamefont {Z.~J.}\ \bibnamefont {Wall}}, \bibinfo {author} {\bibfnamefont {M.~J.}\ \bibnamefont {Boguslawski}}, \bibinfo {author} {\bibfnamefont {M.}~\bibnamefont {Bareian}}, \bibinfo {author} {\bibfnamefont {A.}~\bibnamefont {Derevianko}}, \bibinfo {author} {\bibfnamefont {W.~C.}\ \bibnamefont {Campbell}}, \ and\ \bibinfo {author} {\bibfnamefont {E.~R.}\ \bibnamefont {Hudson}},\ }\href@noop {} {\enquote {\bibinfo {title} {Eliminating qubit type cross-talk in the $\textit{omg}$ protocol},}\ } (\bibinfo {year} {2023}),\ \Eprint {http://arxiv.org/abs/2310.10905} {arXiv:2310.10905 [quant-ph]} \BibitemShut {NoStop}%
\bibitem [{ba1(2015)}]{ba133limit}%
  \BibitemOpen
  \href {https://nuclearsafety.gc.ca/eng/acts-and-regulations/regulatory-documents/published/html/regdoc1-6-1/appendix-r.cfm} {\enquote {\bibinfo {title} {Nuclear substances and radiation devices regulations},}\ } (\bibinfo {year} {2015})\BibitemShut {NoStop}%
\bibitem [{\citenamefont {{Christensen, Justin E.}}(2020)}]{Christensen}%
  \BibitemOpen
  \bibfield  {author} {\bibinfo {author} {\bibnamefont {{Christensen, Justin E.}}},\ }\href@noop {} {\enquote {\bibinfo {title} {High-fidelity operation of a radioactive trapped ion qubit, \textsuperscript{133}ba\textsuperscript{+}},}\ } (\bibinfo {year} {2020})\BibitemShut {NoStop}%
\bibitem [{\citenamefont {White}\ \emph {et~al.}(2022)\citenamefont {White}, \citenamefont {Low}, \citenamefont {de~Sereville}, \citenamefont {Day}, \citenamefont {Greenberg}, \citenamefont {Rademacher},\ and\ \citenamefont {Senko}}]{PhysRevA.105.033102}%
  \BibitemOpen
  \bibfield  {author} {\bibinfo {author} {\bibfnamefont {B.~M.}\ \bibnamefont {White}}, \bibinfo {author} {\bibfnamefont {P.~J.}\ \bibnamefont {Low}}, \bibinfo {author} {\bibfnamefont {Y.}~\bibnamefont {de~Sereville}}, \bibinfo {author} {\bibfnamefont {M.~L.}\ \bibnamefont {Day}}, \bibinfo {author} {\bibfnamefont {N.}~\bibnamefont {Greenberg}}, \bibinfo {author} {\bibfnamefont {R.}~\bibnamefont {Rademacher}}, \ and\ \bibinfo {author} {\bibfnamefont {C.}~\bibnamefont {Senko}},\ }\href {\doibase 10.1103/PhysRevA.105.033102} {\bibfield  {journal} {\bibinfo  {journal} {Phys. Rev. A}\ }\textbf {\bibinfo {volume} {105}},\ \bibinfo {pages} {033102} (\bibinfo {year} {2022})}\BibitemShut {NoStop}%
\bibitem [{\citenamefont {Shi}\ \emph {et~al.}(2023)\citenamefont {Shi}, \citenamefont {Todaro}, \citenamefont {Mintzer}, \citenamefont {Bruzewicz}, \citenamefont {Chiaverini},\ and\ \citenamefont {Chuang}}]{mitsurface}%
  \BibitemOpen
  \bibfield  {author} {\bibinfo {author} {\bibfnamefont {X.}~\bibnamefont {Shi}}, \bibinfo {author} {\bibfnamefont {S.~L.}\ \bibnamefont {Todaro}}, \bibinfo {author} {\bibfnamefont {G.~L.}\ \bibnamefont {Mintzer}}, \bibinfo {author} {\bibfnamefont {C.~D.}\ \bibnamefont {Bruzewicz}}, \bibinfo {author} {\bibfnamefont {J.}~\bibnamefont {Chiaverini}}, \ and\ \bibinfo {author} {\bibfnamefont {I.~L.}\ \bibnamefont {Chuang}},\ }\href {\doibase 10.1063/5.0149778} {\bibfield  {journal} {\bibinfo  {journal} {Applied Physics Letters}\ }\textbf {\bibinfo {volume} {122}},\ \bibinfo {pages} {264002} (\bibinfo {year} {2023})},\ \Eprint {http://arxiv.org/abs/https://pubs.aip.org/aip/apl/article-pdf/doi/10.1063/5.0149778/18024081/264002\_1\_5.0149778.pdf} {https://pubs.aip.org/aip/apl/article-pdf/doi/10.1063/5.0149778/18024081/264002\_1\_5.0149778.pdf} \BibitemShut {NoStop}%
\bibitem [{\citenamefont {et~al}(2020)}]{huculconference}%
  \BibitemOpen
  \bibfield  {author} {\bibinfo {author} {\bibfnamefont {Z.~S.}\ \bibnamefont {et~al}},\ }in\ \href@noop {} {\emph {\bibinfo {booktitle} {51st Annual Meeting of the APS Division of Atomic, Molecular and Optical Physics}}},\ Vol.~\bibinfo {volume} {65},\ \bibinfo {organization} {APS}\ (\bibinfo  {publisher} {American Physical Society},\ \bibinfo {address} {Portland, Oregon},\ \bibinfo {year} {2020})\BibitemShut {NoStop}%
\bibitem [{\citenamefont {et~al}(2022)}]{brennanconference}%
  \BibitemOpen
  \bibfield  {author} {\bibinfo {author} {\bibfnamefont {A.~V.~B.}\ \bibnamefont {et~al}},\ }in\ \href@noop {} {\emph {\bibinfo {booktitle} {53rd Annual Meeting of the APS Division of Atomic, Molecular and Optical Physics}}},\ Vol.~\bibinfo {volume} {67},\ \bibinfo {organization} {APS}\ (\bibinfo  {publisher} {American Physical Society},\ \bibinfo {address} {Orlando, Florida},\ \bibinfo {year} {2022})\BibitemShut {NoStop}%
\bibitem [{\citenamefont {Greenberg}\ \emph {et~al.}(2023)\citenamefont {Greenberg}, \citenamefont {White}, \citenamefont {Low},\ and\ \citenamefont {Senko}}]{greenberg2023trapping}%
  \BibitemOpen
  \bibfield  {author} {\bibinfo {author} {\bibfnamefont {N.}~\bibnamefont {Greenberg}}, \bibinfo {author} {\bibfnamefont {B.~M.}\ \bibnamefont {White}}, \bibinfo {author} {\bibfnamefont {P.~J.}\ \bibnamefont {Low}}, \ and\ \bibinfo {author} {\bibfnamefont {C.}~\bibnamefont {Senko}},\ }\href@noop {} {\enquote {\bibinfo {title} {Trapping $\mathbf{Ba}^+$ with seven-fold enhanced efficiency utilizing an autoionizing resonance},}\ } (\bibinfo {year} {2023}),\ \Eprint {http://arxiv.org/abs/2307.07627} {arXiv:2307.07627 [quant-ph]} \BibitemShut {NoStop}%
\bibitem [{\citenamefont {Rey}\ \emph {et~al.}(2022)\citenamefont {Rey}, \citenamefont {Walter}, \citenamefont {Harrer}, \citenamefont {Perez}, \citenamefont {Chiera}, \citenamefont {Nair}, \citenamefont {Ickler}, \citenamefont {Fuchs}, \citenamefont {Michaud}, \citenamefont {Uttinger}, \citenamefont {Schofield}, \citenamefont {Thijssen}, \citenamefont {Distaso}, \citenamefont {Peukert},\ and\ \citenamefont {Vogel}}]{rey}%
  \BibitemOpen
  \bibfield  {author} {\bibinfo {author} {\bibfnamefont {M.}~\bibnamefont {Rey}}, \bibinfo {author} {\bibfnamefont {J.}~\bibnamefont {Walter}}, \bibinfo {author} {\bibfnamefont {J.}~\bibnamefont {Harrer}}, \bibinfo {author} {\bibfnamefont {C.~M.}\ \bibnamefont {Perez}}, \bibinfo {author} {\bibfnamefont {S.}~\bibnamefont {Chiera}}, \bibinfo {author} {\bibfnamefont {S.}~\bibnamefont {Nair}}, \bibinfo {author} {\bibfnamefont {M.}~\bibnamefont {Ickler}}, \bibinfo {author} {\bibfnamefont {A.}~\bibnamefont {Fuchs}}, \bibinfo {author} {\bibfnamefont {M.}~\bibnamefont {Michaud}}, \bibinfo {author} {\bibfnamefont {M.~J.}\ \bibnamefont {Uttinger}}, \bibinfo {author} {\bibfnamefont {A.~B.}\ \bibnamefont {Schofield}}, \bibinfo {author} {\bibfnamefont {J.~H.~J.}\ \bibnamefont {Thijssen}}, \bibinfo {author} {\bibfnamefont {M.}~\bibnamefont {Distaso}}, \bibinfo {author} {\bibfnamefont {W.}~\bibnamefont {Peukert}}, \ and\ \bibinfo {author} {\bibfnamefont {N.}~\bibnamefont {Vogel}},\ }\href@noop {} {\bibfield  {journal}
  {\bibinfo  {journal} {Nature Communications}\ }\textbf {\bibinfo {volume} {13}},\ \bibinfo {pages} {2840} (\bibinfo {year} {2022})}\BibitemShut {NoStop}%
\bibitem [{\citenamefont {Das}\ \emph {et~al.}(2017)\citenamefont {Das}, \citenamefont {Dey}, \citenamefont {Reddy},\ and\ \citenamefont {Sarma}}]{das}%
  \BibitemOpen
  \bibfield  {author} {\bibinfo {author} {\bibfnamefont {S.}~\bibnamefont {Das}}, \bibinfo {author} {\bibfnamefont {A.}~\bibnamefont {Dey}}, \bibinfo {author} {\bibfnamefont {G.}~\bibnamefont {Reddy}}, \ and\ \bibinfo {author} {\bibfnamefont {D.~D.}\ \bibnamefont {Sarma}},\ }\href {\doibase 10.1021/acs.jpclett.7b01814} {\bibfield  {journal} {\bibinfo  {journal} {The Journal of Physical Chemistry Letters}\ }\textbf {\bibinfo {volume} {8}},\ \bibinfo {pages} {4704} (\bibinfo {year} {2017})},\ \bibinfo {note} {pMID: 28885853},\ \Eprint {http://arxiv.org/abs/https://doi.org/10.1021/acs.jpclett.7b01814} {https://doi.org/10.1021/acs.jpclett.7b01814} \BibitemShut {NoStop}%
\bibitem [{\citenamefont {Hertaeg}\ \emph {et~al.}(2021)\citenamefont {Hertaeg}, \citenamefont {Rees-Zimmerman}, \citenamefont {Tabor}, \citenamefont {Routh},\ and\ \citenamefont {Garnier}}]{HERTAEG202152}%
  \BibitemOpen
  \bibfield  {author} {\bibinfo {author} {\bibfnamefont {M.~J.}\ \bibnamefont {Hertaeg}}, \bibinfo {author} {\bibfnamefont {C.}~\bibnamefont {Rees-Zimmerman}}, \bibinfo {author} {\bibfnamefont {R.~F.}\ \bibnamefont {Tabor}}, \bibinfo {author} {\bibfnamefont {A.~F.}\ \bibnamefont {Routh}}, \ and\ \bibinfo {author} {\bibfnamefont {G.}~\bibnamefont {Garnier}},\ }\href {\doibase https://doi.org/10.1016/j.jcis.2021.01.092} {\bibfield  {journal} {\bibinfo  {journal} {Journal of Colloid and Interface Science}\ }\textbf {\bibinfo {volume} {591}},\ \bibinfo {pages} {52} (\bibinfo {year} {2021})}\BibitemShut {NoStop}%
\bibitem [{\citenamefont {Rösch}(2014)}]{Nuclear}%
  \BibitemOpen
  \bibfield  {author} {\bibinfo {author} {\bibfnamefont {F.}~\bibnamefont {Rösch}},\ }\href@noop {} {\emph {\bibinfo {title} {Nuclear- and Radiochemistry}}}\ (\bibinfo  {publisher} {De Gruyter},\ \bibinfo {year} {2014})\BibitemShut {NoStop}%
\bibitem [{ba1()}]{ba133mass}%
  \BibitemOpen
  \href@noop {} {}\bibinfo {howpublished} {\url{https://pubchem.ncbi.nlm.nih.gov/compound/Barium-133.}},\ \bibinfo {note} {accessed: 2023-11-8}\BibitemShut {NoStop}%
\bibitem [{\citenamefont {Goldstein}\ \emph {et~al.}(2017)\citenamefont {Goldstein}, \citenamefont {Newbury}, \citenamefont {Michael}, \citenamefont {Ritchie}, \citenamefont {Scott},\ and\ \citenamefont {Joy}}]{edss}%
  \BibitemOpen
  \bibfield  {author} {\bibinfo {author} {\bibfnamefont {J.~I.}\ \bibnamefont {Goldstein}}, \bibinfo {author} {\bibfnamefont {D.~E.}\ \bibnamefont {Newbury}}, \bibinfo {author} {\bibfnamefont {J.~R.}\ \bibnamefont {Michael}}, \bibinfo {author} {\bibfnamefont {N.~W.}\ \bibnamefont {Ritchie}}, \bibinfo {author} {\bibfnamefont {J.~H.~J.}\ \bibnamefont {Scott}}, \ and\ \bibinfo {author} {\bibfnamefont {D.~C.}\ \bibnamefont {Joy}},\ }\href@noop {} {\emph {\bibinfo {title} {Scanning Electron Microscopy and X-Ray Microanalysis}}},\ \bibinfo {edition} {4th}\ ed.\ (\bibinfo  {publisher} {Springer New York, NY},\ \bibinfo {year} {2017})\BibitemShut {NoStop}%
\bibitem [{\citenamefont {Revelle}(2020)}]{revelle2020phoenix}%
  \BibitemOpen
  \bibfield  {author} {\bibinfo {author} {\bibfnamefont {M.~C.}\ \bibnamefont {Revelle}},\ }\href@noop {} {\enquote {\bibinfo {title} {Phoenix and peregrine ion traps},}\ } (\bibinfo {year} {2020}),\ \Eprint {http://arxiv.org/abs/2009.02398} {arXiv:2009.02398 [physics.app-ph]} \BibitemShut {NoStop}%
\bibitem [{\citenamefont {Turchette}\ \emph {et~al.}(2000)\citenamefont {Turchette}, \citenamefont {Kielpinski}, \citenamefont {King}, \citenamefont {Leibfried}, \citenamefont {Meekhof}, \citenamefont {Myatt}, \citenamefont {Rowe}, \citenamefont {Sackett}, \citenamefont {Wood}, \citenamefont {Itano}, \citenamefont {Monroe},\ and\ \citenamefont {Wineland}}]{PhysRevA.61.063418}%
  \BibitemOpen
  \bibfield  {author} {\bibinfo {author} {\bibfnamefont {Q.~A.}\ \bibnamefont {Turchette}}, \bibinfo {author} {\bibnamefont {Kielpinski}}, \bibinfo {author} {\bibfnamefont {B.~E.}\ \bibnamefont {King}}, \bibinfo {author} {\bibfnamefont {D.}~\bibnamefont {Leibfried}}, \bibinfo {author} {\bibfnamefont {D.~M.}\ \bibnamefont {Meekhof}}, \bibinfo {author} {\bibfnamefont {C.~J.}\ \bibnamefont {Myatt}}, \bibinfo {author} {\bibfnamefont {M.~A.}\ \bibnamefont {Rowe}}, \bibinfo {author} {\bibfnamefont {C.~A.}\ \bibnamefont {Sackett}}, \bibinfo {author} {\bibfnamefont {C.~S.}\ \bibnamefont {Wood}}, \bibinfo {author} {\bibfnamefont {W.~M.}\ \bibnamefont {Itano}}, \bibinfo {author} {\bibfnamefont {C.}~\bibnamefont {Monroe}}, \ and\ \bibinfo {author} {\bibfnamefont {D.~J.}\ \bibnamefont {Wineland}},\ }\href {\doibase 10.1103/PhysRevA.61.063418} {\bibfield  {journal} {\bibinfo  {journal} {Phys. Rev. A}\ }\textbf {\bibinfo {volume} {61}},\ \bibinfo {pages} {063418} (\bibinfo {year} {2000})}\BibitemShut {NoStop}%
\bibitem [{\citenamefont {Deslauriers}\ \emph {et~al.}(2006)\citenamefont {Deslauriers}, \citenamefont {Olmschenk}, \citenamefont {Stick}, \citenamefont {Hensinger}, \citenamefont {Sterk},\ and\ \citenamefont {Monroe}}]{PhysRevLett.97.103007}%
  \BibitemOpen
  \bibfield  {author} {\bibinfo {author} {\bibfnamefont {L.}~\bibnamefont {Deslauriers}}, \bibinfo {author} {\bibfnamefont {S.}~\bibnamefont {Olmschenk}}, \bibinfo {author} {\bibfnamefont {D.}~\bibnamefont {Stick}}, \bibinfo {author} {\bibfnamefont {W.~K.}\ \bibnamefont {Hensinger}}, \bibinfo {author} {\bibfnamefont {J.}~\bibnamefont {Sterk}}, \ and\ \bibinfo {author} {\bibfnamefont {C.}~\bibnamefont {Monroe}},\ }\href {\doibase 10.1103/PhysRevLett.97.103007} {\bibfield  {journal} {\bibinfo  {journal} {Phys. Rev. Lett.}\ }\textbf {\bibinfo {volume} {97}},\ \bibinfo {pages} {103007} (\bibinfo {year} {2006})}\BibitemShut {NoStop}%
\bibitem [{\citenamefont {Sedlacek}\ \emph {et~al.}(2018)\citenamefont {Sedlacek}, \citenamefont {Greene}, \citenamefont {Stuart}, \citenamefont {McConnell}, \citenamefont {Bruzewicz}, \citenamefont {Sage},\ and\ \citenamefont {Chiaverini}}]{PhysRevA.97.020302}%
  \BibitemOpen
  \bibfield  {author} {\bibinfo {author} {\bibfnamefont {J.~A.}\ \bibnamefont {Sedlacek}}, \bibinfo {author} {\bibfnamefont {A.}~\bibnamefont {Greene}}, \bibinfo {author} {\bibfnamefont {J.}~\bibnamefont {Stuart}}, \bibinfo {author} {\bibfnamefont {R.}~\bibnamefont {McConnell}}, \bibinfo {author} {\bibfnamefont {C.~D.}\ \bibnamefont {Bruzewicz}}, \bibinfo {author} {\bibfnamefont {J.~M.}\ \bibnamefont {Sage}}, \ and\ \bibinfo {author} {\bibfnamefont {J.}~\bibnamefont {Chiaverini}},\ }\href {\doibase 10.1103/PhysRevA.97.020302} {\bibfield  {journal} {\bibinfo  {journal} {Phys. Rev. A}\ }\textbf {\bibinfo {volume} {97}},\ \bibinfo {pages} {020302} (\bibinfo {year} {2018})}\BibitemShut {NoStop}%
\end{thebibliography}%

\end{document}